\newcommand{\myTitle}{Extracting dipole orientations from asymmetric plasmonic nanostructures towards machine-learning-assisted spectropolarimetry}

\documentclass[12pt]{article}
\usepackage[a4paper,
bindingoffset=0.2in,
left=0.634in,
right=0.634in,
top=1in,
bottom=1in,
footskip=.5in]{geometry}
\usepackage{float}
\usepackage{xcolor}
\definecolor{Maroon}{cmyk}{0, 0.87, 0.68, 0.32}
\definecolor{RoyalBlue}{cmyk}{1, 0.50, 0, 0}
\usepackage{epsfig}
\usepackage{caption}
\usepackage{siunitx}
\usepackage{xkeyval}
\usepackage{tocloft}
\usepackage{chemformula}
\usepackage{amsmath}
\usepackage{hyperref}
\usepackage{microtype}
\microtypesetup{activate=false, protrusion=false}

\hypersetup{%
	colorlinks=true, linktocpage=true, pdfstartpage=1, pdfstartview=FitV,%
	breaklinks=true, pdfpagemode=UseNone, pageanchor=true, pdfpagemode=UseOutlines,%
	plainpages=false, bookmarksnumbered, bookmarksopen=true, bookmarksopenlevel=1,%
	hypertexnames=true, %nesting=true,%frenchlinks,%
	urlcolor=Maroon, linkcolor=RoyalBlue, citecolor=RoyalBlue, %pagecolor=RoyalBlue,%
	pdftitle={\myTitle},%
	pdfauthor={\textcopyright\ P. Christian Simo, Eberhard Karls University Tübingen, Institute for Applied Physics},%
	pdfsubject={},%
	pdfkeywords={},%
	pdfcreator={pdfLaTeX},%
	pdfproducer={LaTeX with hyperref and classicthesis}%
}   
\usepackage{graphicx}
\graphicspath{{images/}}

\usepackage{geometry}
\usepackage{mathpazo}
\usepackage[onehalfspacing]{setspace}
\usepackage{scalefnt}
\newcommand\textsmaller[2][0.85]{{\scalefont{#1}#2}}
\usepackage[english]{babel}
\usepackage[utf8]{inputenc}
\usepackage{authblk}

\usepackage{microtype}
\microtypesetup{activate=true, protrusion=false, tracking=true}

\usepackage[style=numeric-comp, backend=bibtex8, bibencoding=ascii, language=auto, natbib=true, sorting=none, maxbibnames=10]{biblatex}
\newcommand{\parens}[1]{\left(#1\right)}
\newcommand{\brackets}[1]{\left[#1\right]}
\newcommand{\braces}[1]{\left\{#1\right\}}

\hypersetup{breaklinks=true}

\usepackage[nolist, withpage]{acronym}
\begin{acronym}
	\acro{ADM}{analytical dipole model}
	\acro{AR}{aspect ratio}
	
	\acro{BEM}{boundary element method}
	\acro{BFP}{back focal plane}
	\acro{BS}{beam splitter}
	
	\acro{CNN}{convolutional neural network}
	\acro{CCD}{charge-coupled device}
	\acro{CV}{cross-validation}
	
	\acro{DDSP}{dipole determination by spectropolarimetry}
	\acro{DFC}{dark-field condenser}
	
	\acro{EBL}{electron beam lithography}
	
	\acro{FWHM}{full width at half maximum}
	
	\acro{HGBR}{histogram gradient boosting regressor}
	
	\acro{ITO}{indium tin oxide}
	\acro{IQR}{interquartile region}
	\acro{IPA}{isopropyl alcohol}
	
	\acro{LP}{linear polarizer}
	\acro{LSPR}{localized surface plasmon resonance}
	\acro{LSP}{localized surface plasmon}
	\acro{LR}{longitudinal resonance}
	
	\acro{MSE}{mean square error}
	\acro{ML}{machine learning}
	
	\acro{NA}{numerical aperture}
	
	\acro{PMMA}{poly(methyl methacrylate) 2041 (447k g/mol)}
	\acro{PGMEA}{propylene glycol methyl ether acetate}
	
	\acro{PSF}{point spread function}
	\acro{PH}{pinhole}
	
	\acro{QNM}{quasi-normal modes}
	
	\acro{RMSE}{root mean square error}
	\acro{aRMSE}{adjusted root mean square error}
	
	\acro{SERS}{surface-enhanced Raman scattering}
	\acro{SVM}{support vector machine}
	\acro{SEM}{scanning electron microscope}
	
	\acro{TE}{transverse electric}
	\acro{TM}{transverse magnetic}
	\acro{TR}{transversal resonance}
	\acro{TL}{tube lens}
	\acro{TMDC}{transition metal dichalcogenide}
	
\end{acronym}

\providecommand{\keywords}[1]
{
	\small	
	\textbf{\textit{Keywords---}} #1
}

\title{\myTitle}

\author[ ]{P. Christian Simo}
\author[ ]{Michaela Zbytovska}
\author[ ]{Annika Mildner}
\author[ ]{Lukas Lang}
\author[ ]{Melanie Sommer}
\author[ ]{Dieter P. Kern}
\author[ ]{Monika Fleischer}

\affil[ ]{\textit{Institute for Applied Physics and Center LISA\textsuperscript{+}, Auf der Morgenstelle 10, 72076 Tübingen}}
\affil[ ]{\textit{christian.simo@uni-tuebingen.de}; \textit{monika.fleischer@uni-tuebingen.de};}
\date{}

\begin{document}
\maketitle

\begin{abstract}
	In this work, nanoparticles with various asymmetries are analyzed for their azimuthal orientations using polarimetric dark-field spectroscopy at different analyzing angles of a linear polarizer. This approach reveals their spectral behavior in terms of electric far-field dipole intensities when modeled with an analytical dipole model. By simultaneously fitting the spectra from a set of analyzer angles, the respective dipole orientations are extracted. In a statistical approach, all non-repeating permutations are further studied with a machine learning algorithm. The resulting azimuthal distribution of dipole orientations coincides well with the geometric orientations derived from simulations and electron microscope images. A histogram gradient boosting regressor evaluates the impact of the measurement setup on the simultaneously fitted sets, linking the weights of the analyzer angles to the asymmetry in the plasmonic systems. This comprehensive spectroscopic method improves the accuracy of dipole orientation measurements and enables modern machine learning models to interpret potentially complex features of nanostructures.  
\end{abstract}

\newpage

\section{Introduction}

The orientation of dipoles, whether transition dipoles in molecules or localized plasmonic modes of nanoparticles, is a key factor that determines radiative properties such as emission polarization, directionality, coupling, and hybridization \cite{novotny2012}. Accurately and efficiently retrieving dipole orientation enables studies in biophysics and physical chemistry for single molecules, aids in designing directional and chiral nanoemitters, and assists in engineering light-matter interfaces for quantum sensing and various applications.

Polarization-resolved detection offers a straightforward way to determine orientation. Techniques like polarimeters and sequential analyzer measurements decompose emitted light into orthogonal polarization components, directly differentiating in-plane versus out-of-plane dipole orientations and rotational diffusion \cite{sickOrientationalImagingSingle2000,liebSinglemoleculeOrientationsDetermined2004,lethiecMeasurementThreeDimensionalDipole2014}. When spectral resolution is added, polarimetric data can distinguish orientation-dependent spectral features, which is especially important in plasmonics. This is particularly relevant for coupled systems and structures with no rotational symmetry, where resonances are sensitive to polarization \cite{yongDeterminingOrientationsOptical2018,zhouExperimentalDeterminationDipole2021}. Polarimetric methods encounter systematic challenges because excitation polarization and anisotropic sample interfaces can bias orientation measurements unless explicitly modeled. The analyzer channel, while essential, decreases the signal and may reduce signal-to-noise ratio if not compensated for with longer measurement times. Additionally, near-field coupling between nanostructures or between molecules and nanostructures can modify effective dipole moments, causing far-field polarimetry to become inaccurate without electrodynamic corrections \cite{debarreQuantitativeDetermination3D2004, zhaoLocalizationAccuracyGold2017}.

Over the past decade, the popularity of deep learning algorithms and convolutional neural net-works has grown significantly. These data-driven techniques have introduced innovative methods for evaluating and understanding phenomena more effectively, also in the field of nano-optics. Convolutional neural networks and similar architectures have been trained on both synthetic and experimental data to determine orientation, wobble, and position from complex images, where manual approaches can be slow or challenging \cite{cid-mejiasDeepLearningApproach2021,hu2020a}. Deep learning offers several practical benefits, including high throughput, robustness to moderate model mismatch, and the capability to perform localization, categorization, and orientation estimation simultaneously in dense fields \cite{debarreQuantitativeDetermination3D2004}. Hybrid approaches that combine physics-based models with physics-trained neural networks or optimized optical components have demonstrated particular promise in surpassing the performance of traditional hardware-software methods \cite{hu2020a}. 

Asymmetry, both geometric and electromagnetic, fundamentally expands the concept of dipole radiation. Chiral nanostructures introduce effects such as circular dichroism and chiral emission patterns, allowing for circularly polarized directional emission and spin-orbit coupling of light \cite{brasseletSinglemoleculeOrientationLocalization2025, kimDipolelikeElectrostaticAsymmetry2018}.

In current literature, the number of theoretical frameworks to describe asymmetric behavior of dipole moment orientations is significant due to their ability to be adapted and modified to fit the experimental procedures. Intrinsic chirality within systems with implications of dipole interferences is proven to be possible with purely electric dipoles \cite{xavierStrongCollectiveChiroptical2026}. An \ac{ADM} is able to emulate the far-field scattering of any dipole moment orientation with respect to the direction of the observer and far-field polarization analysis through a core term describing the dipole interference \cite{mildner2023}. This proved to be essential in researching the chiral far-field originating from achiral structures. \Ac{QNM} theory describes complex plasmonic particles and assemblies with the assumption of non-orthogonal dipole moments \cite{lalanne2018, sauvan2013}. Through the use of a similar interference term as in the \ac{ADM}, the \ac{QNM} theory is able to describe the inherent non-Hermiticity of plasmonics and therefore clearly predict the Purcell factors of nanoemitters in close proximity of plasmonic nanostructures. A prominent theoretical model is the Bohn-Kuhn model, which justifies the optical activity and circular dichroism of molecules and nanostructures through a model that is adjacent to a coupled-harmonic oscillator model \cite{zhaoNOscillatorBornKuhn2024}. The Bohn-Kuhn model is successful in revealing the excitation polarization affinity of the analyzed systems. A recent observable, the polarizability vector, could be theoretically established, which states that the anisotropic properties of a plasmonic system are experimentally accessible \cite{olmos-trigoPolarizabilityVectorPolarimetric2026}. This observable allows us to determine the impact of shell coatings or other attachments to nanoparticles with regard to their spectral shape and Stokes parameter values. Such theoretical models are crucial for understanding the inner workings of asymmetric or chiral particles of various complexities.

Janus particles, featuring a deliberately asymmetric composition or shape, produce spatially varying local fields and anisotropic scattering, which can alter the effective dipole orientation and radiation lobes. Experimental studies on hybrid nanorod-fluorophore systems demonstrate how electrostatic and structural asymmetries result in dipole-like but orientation-dependent emission patterns, which deviate from single-dipole models \cite{kimDipolelikeElectrostaticAsymmetry2018, mingExperimentalEvidencePlasmophores2011}.
 
One approach for extracting the dipole orientation of sub-diffraction limit emitters is through the shapes of their imaged \acp{PSF}. Significant progress has been made with engineered \acp{PSF}, enabling joint \textsmaller{3D} localization and orientation recovery from single \ac{CCD} frames, with polarization-resolved and spectropolarimetric measurements providing robust experimental constraints to attain angular information for various systems. These advancements are greatly accelerated by deep-learning methods that manage complex imaging conditions \cite{hu2020a, backlundSimultaneousAccurateMeasurement2012, stallingaPositionOrientationEstimation2012, tebbenjohannsOptimalOrientationDetection2022}. However, challenges remain, especially with limited photon counts, which are further reduced in multiplexed experiments, creating trade-offs between orientation, localization, and spectral intensity signal-to-noise ratio. When dealing with a molecular dipole coupled to a nanoantenna, most errors stem from the incomplete electrodynamic model, which cannot rely solely on the assumption of a single point dipole and requires effective corrections for coupling \cite{changPlasmonicNanorodAbsorbers2010, zhangQuantumLimitsPrecisely2020}. Machine learning can be used to implement these corrections, but always poses a risk if not applied properly. Overfitting and limited training data can produce highly specific and unreliable models for predictive estimation. Ensuring robustness against experimental drift and unseen data is vital for critical applications \cite{cid-mejiasDeepLearningApproach2021}. 
A second common experimental approach to extracting dipole orientations is polarimetric spectroscopy. This approach exploits the intensity evolution of the resonance peaks with the angular variation of the analyzing polarization filter. Polarimetric spectroscopy has recently developed into a precise quantitative method for determining dipole orientation, capable of providing high-accuracy results and addressing previous issues related to substrate-induced bias \cite{simo2025}. Systematic calibration techniques, such as modeling the entire detection pathway or including the angular collection via the objective’s transmission matrix, can mitigate the effects of detection angle and substrate dependence \cite{dasallasEffectDetectionAngle2022}. Linking dipole sources to helical plasmonic antennas enables chiral and directional emission, establishing a connection between emitter orientation and helicity \cite{kuenChiralDirectionalOptical2024}. Extending orientation analysis into spectral and time domains through transient spectropolarimetry opens new opportunities to simultaneously investigate coupling, chirality, and nonlinear behavior at the nanoscale \cite{wuTheoryDipoleMoment2021}.

The dipole moment orientations of nanostructures are clearly related to the system’s geometry, but in asymmetric particles they are not necessarily obvious. Also, in the aforementioned PSF imaging or polarimetric studies, mostly the dominant dipole moment is reconstructed, leading to larger angular error for nearly degenerate, non-dominant dipole moments for low-aspect-ratio particles. 

The present study, in contrast, uses spectro-polarimetric analysis based on an ADM combined with a \ac{ML} estimator to extract the electric dipole orientation of individual plasmon resonance modes in asymmetric plasmonic systems such as imperfect nanoparticles. Additionally, a systematic variation of the experimental parameters is fed into \ac{ML} algorithms to evaluate the precision of the method as well as the sensitivity of the results to the optical setup, which may help to extract further hidden features.

\section{Experiments}
In our previous work, nearly ideal rotationally symmetric colloidal nanorods with two orthogonal dipolar modes aligned to their long and short axis were spectro-polarimetrically investigated. Their analyzed far-field scattering spectra were fitted with the \ac{ADM}, extracting the in-plane orientations of the longitudinal and transversal resonance mode. To quantify the accuracy of the orientations, the angles of the geometric longitudinal symmetry axis were extracted from \ac{SEM} micrographs \cite{simo2025}. The reliability across a random set of gold nanorods with various aspect ratios was examined, resulting in a standard deviation of only $\pm$2.45°. 

\begin{figure*}[t!]
	\centering
	\includegraphics{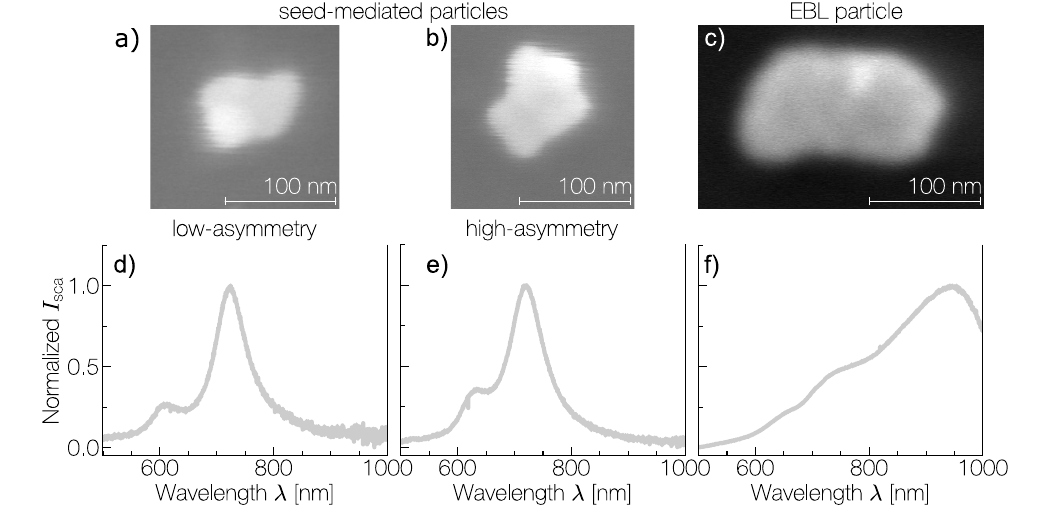}
	\caption{\textbf{(a-c)} \ac{SEM} images of the asymmetric gold nanoparticles studied in this work: \textbf{(a)} low-asymmetry colloidal particle; \textbf{(b)} high-asymmetry colloidal particle, \textbf{(c)} \ac{EBL}-fabricated particle; \textbf{(d-f)} the respective unpolarized dark-field spectra that exhibit (d,e) two or (f) three \ac{LSP} resonances.}\label{fig:specimens}
\end{figure*}

The present work takes a deeper dive, expanding the framework to randomly shaped single nanoparticles as well as probing the influence of the analyzer angles. Three nanostructures with various asymmetries are evaluated polarimetrically, and their analyzed scattering data are subjected to a statistical evaluation, in which the influence of not only the total number of analyzer angles used, but also the analyzer angle combinations, is considered. 

Three plasmonic particles are selected to showcase the statistical insights of this methodology through polarimetric dark-field spectroscopy. The particles contain geometrical asymmetries of various degrees. The particles in our experiment possess plasmonic resonances at distinct wavelengths, which have different dipole moment orientations that we aim to extract, see Figure~\ref{fig:specimens}. Placed on a glass substrate with a layer of indium tin oxide (ITO), two of the particles possess two distinct dipolar \ac{LSP} resonances, while the third particle expresses three \ac{LSP} modes, which will be described in detail later (see Experimental Section \ref{subsec:fabrication} Fabrication). The colloidal particles in Figure~\ref{fig:specimens}a and b are labeled as low-asymmetry and high-asymmetry particles due to their non-uniform geometries observed in the \ac{SEM} after the spectroscopic characterization. They are chemically synthesized through seed-mediated growth by variation of the recipe provided in reference \cite{scarabelliTipsTricksPractical2015}. The third particle (see Figure~\ref{fig:specimens}c) is fabricated through conventional \ac{EBL}. The \ac{EBL}-fabricated particle is designed such that its longitudinal axis is aligned to the 0° polar coordinate of the optical marker system. This particle’s asymmetry is enforced by overdevelopment, leading to a larger size, which can express three \ac{LSP} modes, and an imperfect outline. The plasmonic structures need to undergo optical characterization for their spectral and spatial correlated analysis in the optical setup and \ac{SEM}; hence, a common on-sample coordinate system is established (see Supporting Information). All three plasmonic specimens are spectroscopically characterized with broadband white-light dark-field transmission scattering (see Supporting information, Optical setup and alignment), specifically unpolarized dark-field spectroscopy (cf. Figure \ref{fig:specimens}d-f) and spectro-polarimetry (see Experimental Section \ref{subsec:measurements} Polarimetric measurements).

It is well known that the dipole moments of the respective LSP modes within an ideal nanorod follow the longitudinal and transverse geometric symmetry axes \cite{odellFourOneParallel2026}. As the geometric asymmetry of the nanorod increases, deducing the precise orientation becomes less straightforward. The in-plane orientations of the dipole moments can be roughly estimated as being aligned to the axes of highest symmetry through SEM observations after the spectro-polarimetric experiments \cite{simo2025}. By modeling the particles and simulating them, we also can extract more reliable orientations for the dipole moments through electrodynamic calculations of the surface charge densities. These are used as control angles for each plasmonic system, to qualitatively verify our findings. The control angles will be discussed below.

First, some background information on dipole orientation, far-field emission, and its polarization analysis is provided. Initially, a dipole is induced in a nano-particle by an incident plane wave from above. An unpolarized scattering spectrum is recorded for the extraction of the dipolar intensities and resonance positions for the calculation of the phase responses. Dipolar intensities and phase responses are relevant as fitting parameters for the \ac{ADM}. In the next step, the linearly polarized light is locked at a specific angle by a polarizing filter. The induced dipole then radiates into the far field, which is collected by an objective in the negative \textit{z}-direction. The collected light passes through a second polarizing filter at a variable angle to analyze the electric far-field radiation as a function of the analyzer angle $\Theta$. This results in a set of spectra, one for each analyzing polarizer angle, which are fitted with the \ac{ADM} established in our previous work \cite{simo2025} (see Equation \ref{eq:ADM} in the Experimental Section \ref{subsec:ADM}) that emulates the electric far-field scattering intensity of dipole moments. The precondition for the polarimetric analysis is that the spectra exhibit at least two separate peaks with spectral overlap. The orientations of the respective dipole moments are the fitting parameters to be extracted by this method. The precision of the dipole moment orientation extraction increases with the number of spectra to be simultaneously fitted, while a minimum of three spectra need to be fitted simultaneously to provide sufficient parameter stability and reliable relative orientations \cite{simo2025}. The low errors can be additionally optimized by means of \ac{ML}. \\

To study the accuracy of the extracted dipole moment orientation for asymmetric nanoparticles, we vary the number $m$ of analyzer angles $\Theta$ and observe their statistical influence. Sets of spectra are recorded at fixed intervals of 15° from 0° to 165° for the analyzer angles $\Theta$ (angles modulo 180° are taken as equivalent). While fitting all the resulting spectra simultaneously yields the lowest error, the minimum number of fits for a given acceptable precision is targeted for fast measurements, thus the statistical convergence of the technique needs to be evaluated. Therefore, given the $n$\,=\,12 total analyzer angles, they can be bundled into sets of one to eleven angles, $\{m \in \mathbb{N} \mid 1 \le m \le 11\}$, resulting in specific numbers of possible analyzer angle permutations $C$ derived from the binomial  $C_m^n= \binom{n}{m}=\frac{n!}{\parens{n-m}!m!}$, meaning $C_{1\le m < n}^{n=12} = \brackets{12, 66, 220, 495, 792, 924, 792, 495, 220, 66, 12}$. In the case $m$\,=\,$n$ all available spectra are used, which yields one single combination. For each permutation, the analyzer angles are sorted in ascending order. Each permutation within a set $m$ of analyzer angles then delivers a set of values for the fitted dipole moment orientations, which yield a mean and standard deviation per \ac{LSP} mode $k$. Through the standard deviation $\sigma_m$, the angular spread of this method’s accuracy can be estimated. Additionally, the geometrical highest symmetry axes are extracted as $\gamma_\mathrm{long}$ and $\gamma_\mathrm{short}$ from the \textsmaller{2D} contours of the separate particles. For this purpose, the image of a particle’s contour area and its horizontally mirrored equivalent are overlaid and rotated in opposite directions. The areal overlap is plotted against the angle between the mirror axis and the rotated image. The local maxima of the overlap indicate the angles between the mirror axis and the axes of highest symmetry. A detailed procedure is provided in the Supplementary Information of reference \cite{simo2025}. Additionally, the orientations of the electrical dipoles are calculated from numerical simulations of the induced surface charge densities under illumination for comparison, see Section \ref{subsec:BEM}.

\section{Results}

\begin{figure*}[t!]
	\centering
	\includegraphics[width=\textwidth]{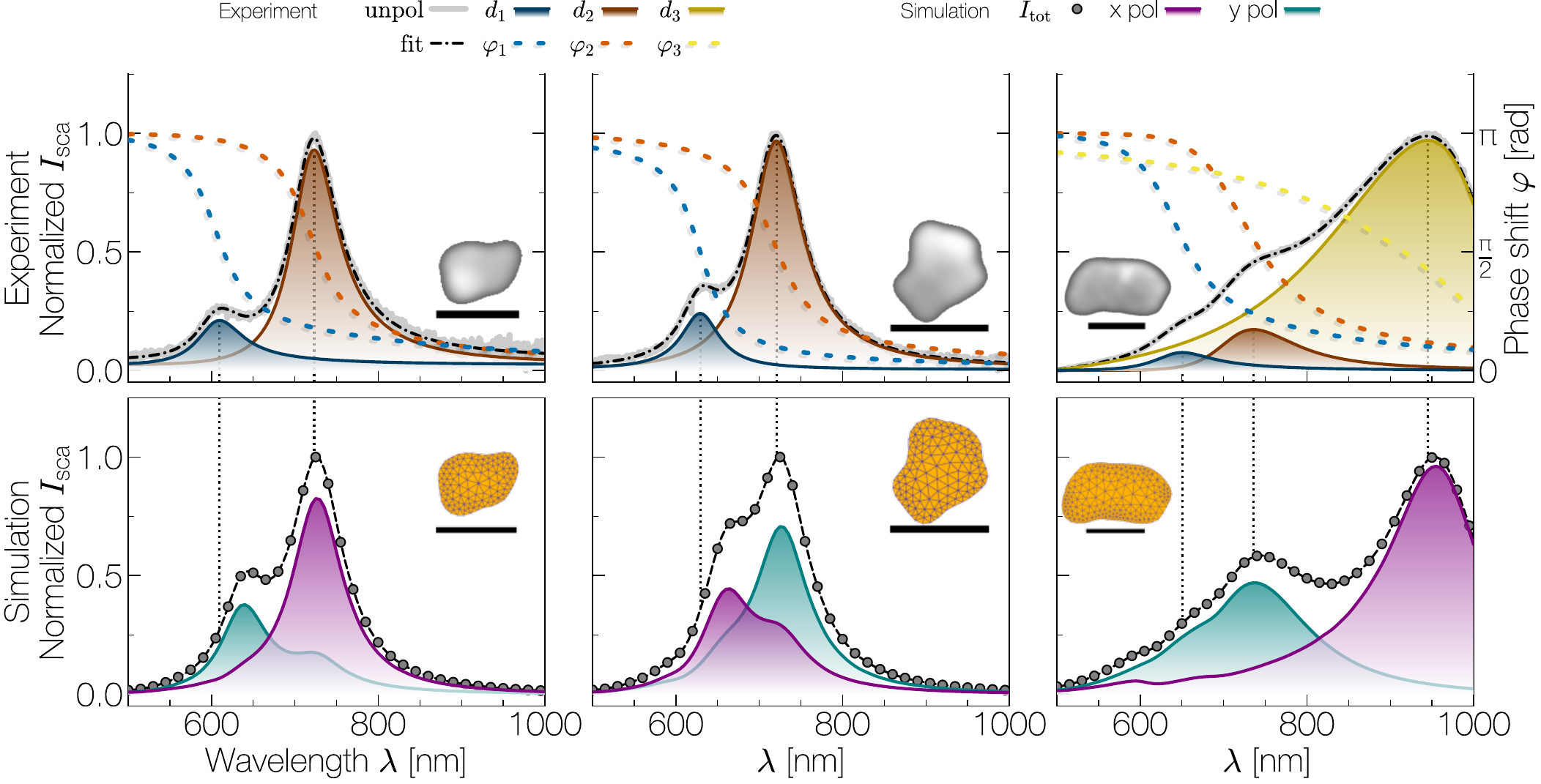}
	\caption{\textbf{(top)} Normalized spectra for the corresponding particles under unpolarized excitation, with the fitted Lorentzian magnitudes and phase shifts. The respective particles are presented as inlays (scale bars: 100 nm). \textbf{(bottom)} The simulated spectra under \textit{x}- and \textit{y}-polarization of the resulting \textsmaller{3D} models. The total intensity is the sum of the simulated spectra, which is used to normalize the data. The top-down views of the extruded and meshed models are also presented. The dotted vertical lines indicate the \ac{LSP} modes from the experiment.}\label{fig:exp_vs_sim}
\end{figure*}

The three asymmetric plasmonic systems given above are analyzed in terms of dipole moment orientation $\gamma_\mathrm{k}$ compared to their geometrical symmetry axis orientations $\gamma_\mathrm{long}$ and $\gamma_\mathrm{short}$ and simulated dipole moment orientations $\gamma_\mathrm{k,sim}$. The normalized and fitted dark-field transmission scattering spectra and the respective simulated scattering intensities are displayed in Figure~\ref{fig:exp_vs_sim}. The contours of the particles are extracted from \ac{SEM} micrographs. The height profiles of the models are adjusted by imitating the grayscale values within the particle contours. The scaling factors or the \textit{z}-coordinate are adapted such that a best fit is achieved between the scattering spectra and corresponding \ac{BEM} simulations (Figure~\ref{fig:exp_vs_sim}, bottom row), see also Section \ref{subsec:BEM}. The dipole magnitudes $d_\mathrm{k}$ and phase responses $\varphi_\mathrm{k}$ are extracted from these unpolarized measurements. The convention of numbering the dipole moments $k$ of a system is in ascending order, from shorter to longer wavelengths. 

\begin{figure*}[t!]
	\centering
	\includegraphics[width=\textwidth]{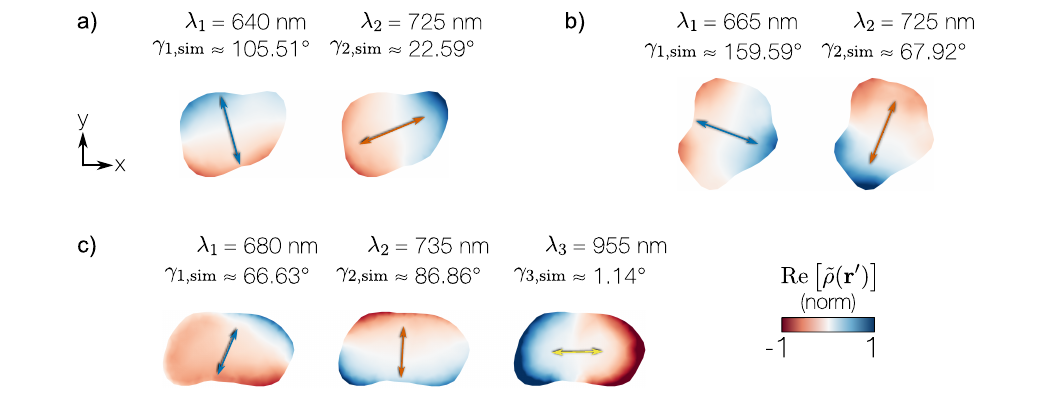}
	\caption{Top view of the surface charge density $\rho\parens{\mathbf{r}'}$ distributions of the three systems at resonance wavelengths $\lambda_\mathrm{k}$ taken from the respective simulated spectra shown in Figure~\ref{fig:exp_vs_sim} (d, e, f). While the \textbf{(a, b)} colloidal particles present two dipolar modes, the \textbf{(c)} \ac{EBL}-fabricated nanostructure shows three modes, from which the first one at $\lambda_1$\,=\,680\,nm is a mixed quadrupolar-dipolar distribution (see Supporting Information Figure S3). Their dipole moments are displayed as arrows with their orientation denoted by $\gamma_\mathrm{k,sim}$. }\label{fig:sim_dipole_moments}
\end{figure*}

In Figure~\ref{fig:sim_dipole_moments}, the simulated models with their respective surface charge densities at the LSP resonance wavelengths $\lambda_\mathrm{k}$ and the resulting dipole moment orientations are illustrated. The coincidence of the dipole moment orientations with some of the particles’ morphological features is apparent. For particularly asymmetric or even chiral particles (see Figure~\ref{fig:sim_dipole_moments}c), the non-orthogonality of the dipole moments is evident, which also leads to stronger dipole coupling and consequently to circular dichroism \cite{fanChiralNanocrystalsPlasmonic2012, melendezBreakingPlasmonicSymmetry2020}. The dipole moments shown in Figure~\ref{fig:sim_dipole_moments}c at 680\,nm and 735\,nm are classified as being quasi-orthogonal to the longest geometric axis of the specimen. Although the particle’s charge distribution at 680\,nm shows quadrupolar characteristics, due to the asymmetry of the quadrupole a dipolar contribution can still be extracted, whereas the remaining symmetric quadrupolar contribution is expected to constitute a dark mode with negligible far-field emission \cite{liuExcitationDarkPlasmons2009}. A supporting argument for this assumption is the scalar decomposition of the surface charge density distributions, which delivers the multipole weights at the LSP mode resonance wavelengths (see Supporting Information, Figure S3) \cite{griffiths2013}. These weights reveal a high dipolar contribution with an added in-phase \textit{x-y} quadrupole. Reconstructing the charge density distribution with the spherical harmonics and their respective multipole weights reveals a distribution that is very close to Figure~\ref{fig:sim_dipole_moments}c (at $\lambda_1$\,=\,680\,nm). None of the displayed dipole moments possess a significant out-of-plane component in the simulated models.

The polarimetric measurements, recorded at a static linear polarization angle $\Phi$\,=\,45°, and analyzer angle $\Theta$ increments of 15° from 0° to 165°, are fitted with the respective \ac{ADM} for two dipoles (see Equation~\ref{eq:ADM}) for the colloidal plasmonic particles and three dipoles (see Supporting Information, Equation S1) for the \ac{EBL}-fabricated particle. Since polarization and analyzing angles $\Phi$ and $\Theta$, as well as the dipole magnitudes $d_\mathrm{k}$ and phase responses $\varphi_\mathrm{k}$ are known, the respective dipole moment orientations, $\gamma_\mathrm{k}$, can be extracted. 

\begin{figure*}[t!]
	\centering
	\includegraphics[width=\textwidth]{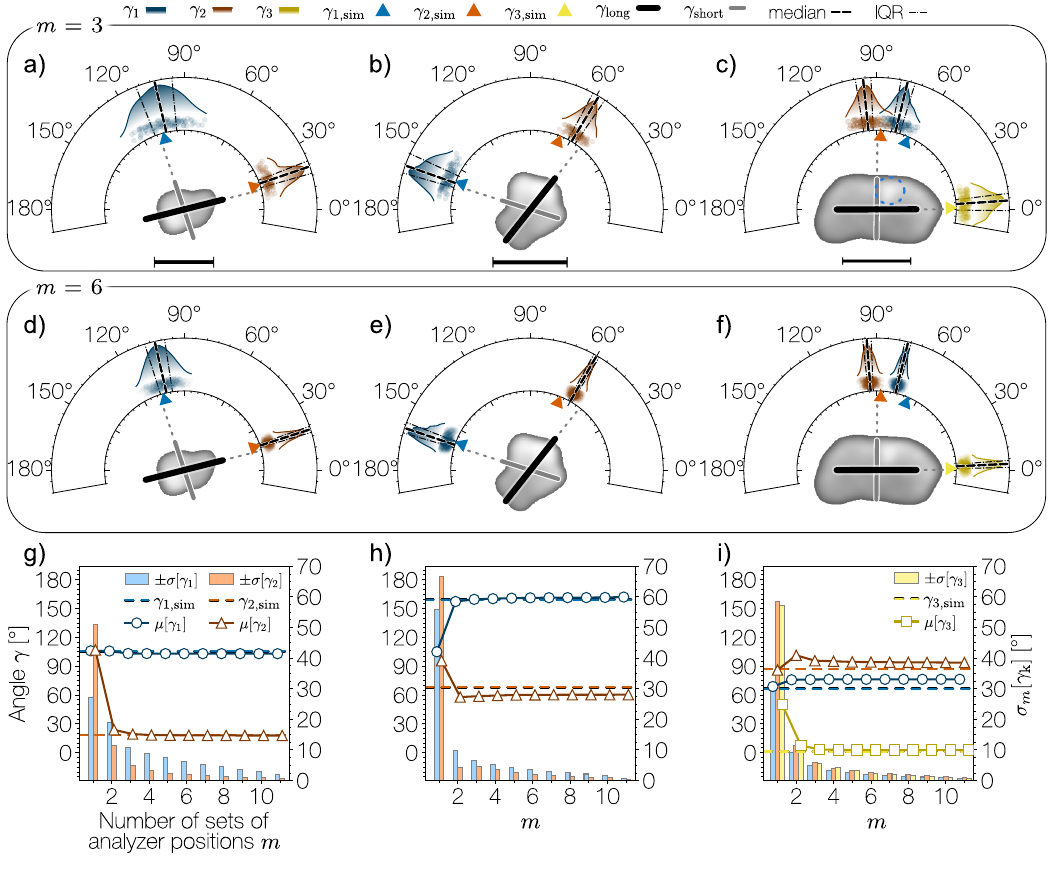}
	\caption{The three asymmetric plasmonic particles as presented in Figure~\ref{fig:exp_vs_sim} and the statistics of the evaluated dipole moment orientations $\gamma_\mathrm{k}$ extracted from all the possible filtered combinations $\tilde{C}$ in the set of $m$\,=\,3 or 6 analyzer angles. The low-asymmetry particle \textbf{(a, d)} and the high-asymmetry particle \textbf{(b, e)} express two distinct dipole moment orientations, while the larger \ac{EBL} particle \textbf{(c, f)} displays three, which mostly show good overlap with the angles of the simulated dipole moment orientations $\gamma_\mathrm{k,sim}$ (internal rim triangles) and the long and short highest symmetry axes at angles $\gamma_\mathrm{long}$ (black bar) and $\gamma_\mathrm{short}$ (gray bar), respectively. The \ac{EBL} particle’s vertical protrusion is marked by the blue dashed ellipse. The \ac{SEM} contours include scale bars of 100\,nm length. The azimuthal spread of the $\gamma_\mathrm{k}$ data is displayed in both a scatterplot and a half-violin plot, which are characterized by the median and the \ac{IQR} of the data. \textbf{(g-i)} The collected dipole orientation angles $\gamma_\mathrm{k}$ for each \ac{LSP} mode for the colloidal nanostructures with \textbf{(a)} low asymmetry, \textbf{(b)} high asymmetry, and \textbf{(c)} the larger \ac{EBL}-fabricated nanostructure for all set numbers $1\le m \le 11$. The respective simulated dipole orientations $\gamma_\mathrm{k,sim}$ (dashed lines) are included with their mean values $\mu\brackets{\gamma_\mathrm{k}}$ and standard deviation values $\sigma_m\brackets{\gamma_\mathrm{k}}$.}\label{fig:spread_fits}
\end{figure*}

From the analyzer angle $\Theta$ variation, a total of 12 different polarimetric spectra are obtained for each plasmonic particle. Computationally, sets of $m$ spectra that are recorded at the respective analyzer angles are fitted simultaneously. Iterating through all the possible permutations $C_m^{n=11}$ with $m\in\braces{1 \le m \le 11}$ delivers a pool of slightly varying values for the in-plane dipole orientations $\gamma_\mathrm{k}$ of each \ac{LSP} mode $k$. An example for a specific permutation procedure is portrayed in the Supporting Information, Figure S2. For conciseness, rather than constantly referring to the polarimetric measurement done at a specific analyzer angle, we just refer to the \textit{analyzer angle} or \textit{analyzer} from here on out. All the extracted angles are collected and statistically evaluated. In Figure~\ref{fig:spread_fits}, the extracted dipole moment orientations $\gamma_\mathrm{k}$ and their simulated in-plane dipole moment orientations $\gamma_\mathrm{k,sim}$ (see Experimental Section \ref{subsec:BEM}) are illustrated together with the influence of the number of simultaneously fitted spectra $m$. The orientation $\gamma_\mathrm{k,sim}$ is calculated from the surface charge densities at the respective \ac{LSP} mode resonance wavelength $\lambda_\mathrm{k}$ as displayed in Figure~\ref{fig:sim_dipole_moments}. For the two colloidal particles (Figure~\ref{fig:spread_fits}a and b), the dipole orientations for the longitudinal and transversal modes follow their geometric symmetry axes at orientations $\gamma_\mathrm{long}$ and $\gamma_\mathrm{short}$. They are oriented intuitively for the low-asymmetry particle but for the high-asymmetry particle they are less obvious due to its increased non-uniformity (see grayscale inset of Figure~\ref{fig:spread_fits}b). As discussed, the \ac{EBL}-fabricated particle (Figure~\ref{fig:spread_fits}c) expresses a third mode at lower wavelengths, which is a mix between a dipolar and quadrupolar surface charge distribution (see Supporting Information, Section 5). 

The collected azimuthal angles of the three asymmetric plasmonic particles from each combination of polarimetric spectra for $m$\,=\,3 (Figure~\ref{fig:spread_fits}a-c) and $m$\,=\,6 (Figure~\ref{fig:spread_fits}d-f) are presented alongside the values of the simulated dipole moment orientations $\gamma_\mathrm{k,sim}$ and the angles $\gamma_\mathrm{long}$ and $\gamma_\mathrm{short}$ of their geometric highest symmetry axes. A key observation here is the azimuthal spread of the fitted dipole orientations $\gamma_\mathrm{k}$, showcased as both scattered points and the resulting half-violin plot. Outliers are excluded from the population, resulting in a filtered population $\tilde{C}$ of orientations from the respective permutation. As the set size $m$ increases, the spread narrows and aligns very well with the evaluated simulation and geometric highest symmetry axes. The \acf{IQR} azimuthal spread includes 50\% of the dipole moment orientations evaluated throughout all the possible analyzer angle permutations $\tilde{C}$ of the respective set $m$. An observed slight systematic offset of some $\gamma_\mathrm{k,sim}$ to the extracted $\gamma_\mathrm{k}$ might originate from the under- or overestimation of the correct height profile of the simulation model (see Section \ref{subsec:BEM}). For the low-asymmetry particle (Figure~\ref{fig:spread_fits}a), this deviation is low, but for the high-asymmetry and \ac{EBL}-fabricated particles it appears to occur in the general azimuthal directions of the highest grayscale values. The good coincidence between the angles $\gamma_\mathrm{k,sim}$ and the individual medians of the $\gamma_\mathrm{k}$ populations makes it evident that the electrodynamic calculations profit from the height profile of the model. 

Focusing on Figure~\ref{fig:spread_fits}a and d, the \ac{IQR} of the longitudinal mode dipole moment orientations $\gamma_2$ decreases from 6.7° to 2.6° as the analyzer set grows from $m$\,=\,3 to $m$\,=\,6. Conversely, the transversal mode orientation $\gamma_1$ shows \ac{IQR} spreads from 16° to 8.8° with increasing $m$. Although the number of permutations is also significantly higher for $m$\,=\,6, we believe the main reason to be the use of six polarimetric spectra for the simultaneous fitting procedure. The difference between the statistics of $\gamma_1$ and $\gamma_2$ is influenced both by the lower signal-to-noise ratio of the transversal mode and the particle’s geometric features. Although this particle does not show strong asymmetry, some variation in particle morphology can be recognized in the transversal orientation. The half-violin spread in Figure~\ref{fig:spread_fits}a hints at two broad regions that overlap, which may indicate two nearly degenerate transversal modes. The diminishing \ac{IQR} azimuthal spread with increasing analyzer set number $m$ increases the confidence of the values achieved with this method, as reflected in the low standard deviation $\sigma_m\brackets{\gamma_\mathrm{k}}$ (see Figure~\ref{fig:spread_fits}g). For higher set numbers $m >$\,2, one observes a clear convergence towards the simulated mode orientations with decreasing $\sigma_m\brackets{\gamma_\mathrm{k}}$, stating the increased precision of the delivered values within the set. Standard deviations of below 2.5° down to sub-degree values can be achieved when most of the polarimetric spectra are used.

Continuing with the high-asymmetry particle (see Figure~\ref{fig:spread_fits}b and e), we note that the statistical spread of the dipole moment orientations $\gamma_2$ lies between the angles of the longitudinal geometric axis $\gamma_\mathrm{long}$ and the simulation $\gamma_\mathrm{2,sim}$. The transversal mode orientation $\gamma_1$, however, overlaps extraordinarily well with the simulated equivalent. Similarly to the previous particle, the half-violin spreads show dipole moment orientations with a clear median peak with an additional shoulder. The azimuthal spread of the \ac{IQR} decreases in a similar fashion with increasing $m$. Although Figure~\ref{fig:spread_fits}h shows low standard deviations for higher $m$, the convergence for the transversal orientation $\gamma_1$ is stronger than that of the longitudinal dipole orientation $\gamma_2$. This may originate from the approximated height profile of the simulation model, as the longitudinal orientations coincide with the highest protrusions of the \textsmaller{3D} simulation model profile. 

Lastly, the \ac{EBL}-fabricated system displayed in Figure~\ref{fig:spread_fits}c and f proves the \ac{ADM}’s capabilities of handling multiple dipole moment orientations. The 0° tilt in the \textit{x-y} plane intended by alignment is confirmed with $\gamma_\mathrm{long}$, but the corresponding dipolar mode orientation deviates by a few degrees. The two transversal modes appear to have two distinct orientations, which may be explained by the asymmetric shapes along the transversal axis. As suggested by the simulated surface charge density distributions in Figure~\ref{fig:sim_dipole_moments}c, the particle’s asymmetric topology and single protrusion positioned at around 75° on the upper half of the particle heavily influence the first transversal orientation $\gamma_1$ to experience a clockwise shift from a 90° orientation towards 75° (see blue dashed circle in Figure~\ref{fig:spread_fits}c). As mentioned, the out-of-plane components of the simulated dipole moments are negligible, with angular deviations from the \textit{x-y} plane in the range of 1° or lower. The slanted edge on the left side may have additionally influenced the asymmetric distribution of the mixed dipolar and quadrupolar in-plane mode $\gamma_1$ that is evident in the scalar decomposition and recomposition of the surface charge densities (see Supporting Information, Section 5). 

A non-uniform relief on the surface, resulting from polycrystallinity in the fabrication processes, clearly affects the evolution of the real dipole moment in the plasmonic system. This is evident when noting that the simulated scattering spectra of the \textsmaller{3D} models do not fully match the experimental ones, particularly for the lower wavelength modes (see Figure~\ref{fig:exp_vs_sim}). Conversely, it is essential to highlight that the all-spectroscopic method, which provides the statistical data, considers the complete picture of the plasmonic system. It should be noted that the collection efficiency of the objective’s \ac{NA} only affects intensity, but not the relevant spectral shape, of the in-plane dipole modes. This makes the method valuable for experiments that are limited to spectroscopic analysis with low photon counts. Estimating the long and short geometrical axes of highest symmetry can be done through the \textsmaller{2D} particle projections in SEM observations. For shapes of C\textsubscript{2} symmetry, this is a straightforward procedure, but for asymmetric shapes, the extraction of the axes is a challenge. This is a secondary benefit of the spectroscopic method. 

Revisiting the mean and standard deviations delivered from this evaluation per set number $m$, displayed in Figure~\ref{fig:spread_fits}g-i, further insights can be gained. A convergence towards the simulated angles is observed around the value of $m$\,=\,3 for most resonance modes, supporting the initial claim of angle stability with at least three analyzer angles \cite{simo2025}. Deviations from the simulated orientations $\gamma_\mathrm{k,sim}$ aside, the standard deviation decreases to sub-degree values, stating the quality of the evaluated mean angles. The large number of possible permutations enables an extensive data population to be evaluated with common statistical approaches as well as with \ac{ML} algorithm models. For $m >$ 6 the number of permutations decreases, but the standard deviation continues to decrease. The main factor in the lower standard deviation thus appears to be the higher number of polarimetric spectra used for the simultaneous fitting of the dipole moment orientations $\gamma_\mathrm{k}$. Simultaneously fitting the full twelve spectra is thus expected to reveal the best-fitted azimuthal orientations of the dipole moments for each particle (see Table~\ref{tab:best_fits}).

\begin{table}
	\centering
	\caption{ADM-extracted orientations $\gamma_\mathrm{k}$ for the three particles obtained by simultaneously fitting all twelve available polarimetric spectra.}\label{tab:best_fits}
	\small{
	\begin{tabular}{r|c c c}
		Particle & $\gamma_\mathrm{1,best}$ & $\gamma_\mathrm{2,best}$ & $\gamma_\mathrm{3,best}$\\[0.5em]
		\hline
		Low-asymmetry & 102.89° $\pm$ 0.63° & 17.73° $\pm$ 0.24° &  -  \\[0.5em]
		High-asymmetry & 161.47° $\pm$ 0.57° & 60.24° $\pm$ 0.14° & - \\[0.5em]
		EBL-fabricated & 76.28° $\pm$ 0.40° & 93.66° $\pm$ 0.18° & 2.59° ± 0.07°\\ [0.5em]
	\end{tabular}
	}
\end{table}

\begin{figure*}[t!]
	\centering
	\includegraphics[width=\textwidth]{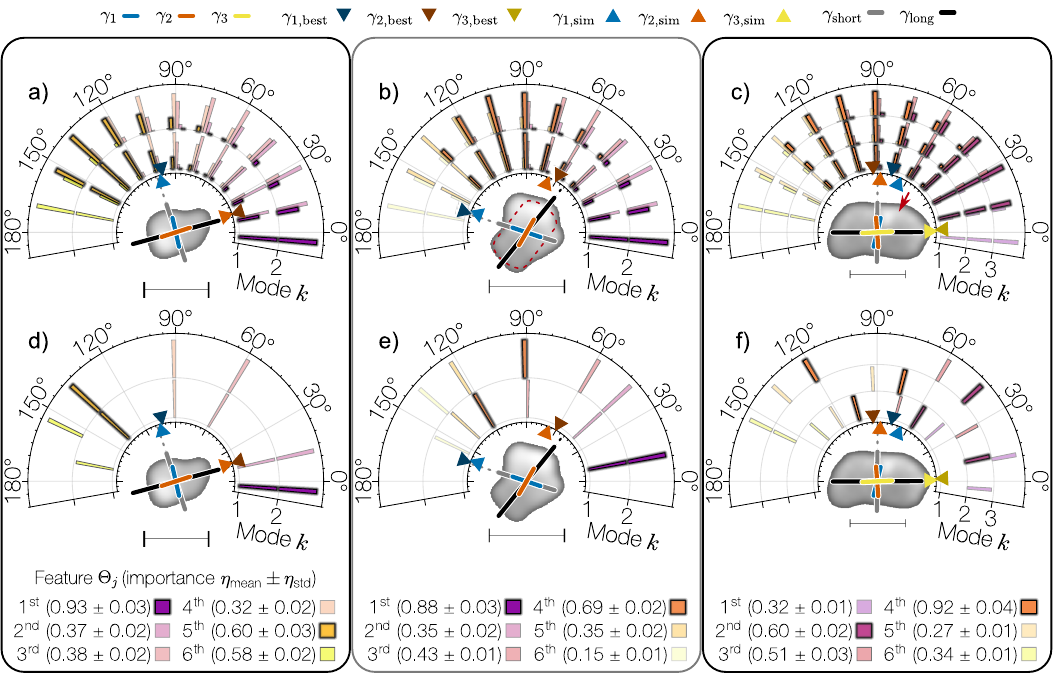}
	\caption{The number of occurrences of each analyzer angle within the permutations (histogram bars) \textbf{(a, b, c)} for all the filtered analyzer angle permutations $\tilde{C}$ within the set $m$\,=\,6. \textbf{(d, e, f)} The analyzer angle permutation which delivers the dipole orientation closest to the angle $\gamma_\mathrm{k,best}$ (inner triangles). The results are presented column-wise for each shown plasmonic system (scale bars 100 nm). The simulated dipole orientations $\gamma_\mathrm{k,sim}$ (outer triangles) and the \ac{SEM}-extracted geometrical axes $\gamma_\mathrm{long}$ (black bar) and $\gamma_\mathrm{short}$ (gray bar) are also depicted. With the set number $m$\,=\,6, the six analyzer angles are evaluated concerning their frequency of appearance within the discretized 15° steps. Their mean importance and standard deviation $\eta_\mathrm{mean}$ and $\eta_\mathrm{std}$ are presented in parentheses and are linked to the respective analyzer angle by color and transparency. The two highest $\eta_\mathrm{mean}$ and their respective histogram bars are emphasized with a dark shadowing rim.}\label{fig:ML_stats}
\end{figure*}

As the set number $m$\,=\,6 delivers a large population with $\tilde{C}$\,$\approx$\,900 separate azimuthal orientation values, we can implement an \ac{ML} algorithm to fit the data for the respective analyzer permutations. The model of choice is a \ac{HGBR} that is further described in Methods. The trained and validated \ac{HGBR} model can be used as an estimator that delivers dipole orientations based on the permutations of analyzer angles, as in the dataset it was trained and tested with. It can predict how the yielded dipole orientation behaves with variation of the single analyzer angle features $\Theta_\mathrm{j}$ and report the degradation of the prediction. This is commonly known as the so-called importance value $\eta$ of a feature $\Theta_\mathrm{j}$. The mean importance $\eta_\mathrm{mean}$ quantifies the average degradation in performance caused by randomizing the relationship between that feature and the target variable, thereby indicating its overall contribution to predictive accuracy. A high $\eta_\mathrm{mean}$ for a feature signifies a strong dependency of the model’s performance on the respective feature. Conversely, low and near-zero $\eta_\mathrm{mean}$ values have little influence, while negative values represent noise or redundancy. The standard deviation of the importance $\eta_\mathrm{std}$ shows the variability of this degradation effect if repeated for a set of permutations. At this point, we observe 100 permutations that deliver predictions closest to the angles $\gamma_\mathrm{k,best}$. This ensures the validity of $\eta_\mathrm{mean}$ and $\eta_\mathrm{std}$ of our \ac{HGBR} model. These can be observed in Figure~\ref{fig:ML_stats}a-c, alongside the occurrence how often each angle appears as the  $j$-th feature $\Theta_\mathrm{j}$ within the set [$\Theta_1$, $\Theta_2$, $\Theta_3$, $\Theta_4$, $\Theta_5$, $\Theta_6$], depicted as a histogram within each 15° step. A color represents each analyzer angle feature, while the transparency of the color is correlated to the analyzer angle’s respective $\eta_\mathrm{mean}$. The two features of the highest importance are indicated by a shadowing rim along their corresponding histogram bars. Alongside the analyzer angle histograms, the best-fit dipole orientations are included together with the correlated angles of the geometric axes and the simulated dipole moment orientations. Figure~\ref{fig:ML_stats}d-f shows the analyzer angle permutations that deliver the $\gamma_\mathrm{k}$ values closest to $\gamma_\mathrm{k,best}$ for each \ac{LSP} mode $k$. In Figure~\ref{fig:ML_stats}a, the $\eta_\mathrm{mean}$ for both dipoles is highest for the first analyzer angle feature, suggesting the low-asymmetric nanoparticle’s longitudinal orientation is predominantly decisive for the data variation in the dataset of the dipole orientations. The asymmetry along the longitudinal axis is located around the same 0°-60° region as the analyzer feature $\Theta_1$, which has the highest importance value. The second-highest importance value is evaluated for $\Theta_5$, which is within the rough azimuthal region of 90°-150° and is likely related to the variation of the model-predicted orientations of the transversal dipole moment. Observing Figure~\ref{fig:ML_stats}d, the ideal permutation shows that the 1\textsuperscript{st} analyzer is indeed in the 0° position, while the other analyzers are represented by a widely spaced set of analyzers that is almost identical for both dipole orientations (see Figure~\ref{fig:ML_stats}d). This supports the assumption that the robustness of the evaluated dipole orientation benefits from a broad angular spread of the analyzer angles. A similar trend can be seen in the highly asymmetric nanoparticle in Figure~\ref{fig:ML_stats}b, where the two analyzer features of higher importance ($\Theta_1$ and $\Theta_4$) are located between the longitudinal and transversal dipoles of the particle, in the azimuthal ranges of 0°-30° and 60°-120°. Using the longitudinal orientation $\gamma_\mathrm{long}$, we can outline a rectangular silhouette (see red dashed contour on inset in Figure~\ref{fig:ML_stats}b), to fit in the \textit{x-y} projection of the nanoparticle. The protrusions beyond the silhouette roughly coincide with the azimuthal ranges of the analyzer features of higher importance. The variation of these specific analyzer features $\Theta_\mathrm{j}$ has a clear degradational effect on the model-predicted orientation of the respective permutation. This is a hint that the importance value $\eta_\mathrm{j,mean}$ of the respective analyzer feature $\Theta_\mathrm{j}$ appears to be sensitive towards the orientation of the geometric features that add to a particle's asymmetry, in which case it could serve as a method to reveal hidden features of the sub-diffraction morphology.

Different ideal permutations for different \ac{LSP} modes are even more evident for the \ac{EBL}-fabricated particle. In Figure~\ref{fig:ML_stats}c, the analyzer feature with the highest importance value is located at the transversal orientation of the nanostructure. Observing the grayscale distribution in the micrograph, one can see higher values along the upper side of the particle, from 30° to 150°, with a prominent protrusion at roughly 75°, which coincides with the angular spread of the analyzers of higher importance.  In the angular spread of the ideal permutation of analyzer angles in Figure~\ref{fig:ML_stats}f, the highest-importance analyzers are again located in the directions of asymmetric features. In this scenario, however, the high-importance features $\Theta_2$ and $\Theta_4$ appear to be correlated to the vertical asymmetry at the northern rim. 

It should be noted that as a precondition, sufficient intensity and dipolar resonances with sufficient spectral overlap are necessary to use the \ac{ADM}’s full potential in the fitting process. In-plane dipolar resonances and their dipole moment orientations can then be extracted very well, even when particles have an asymmetric geometry. The results indicate that the high-importance analyzer features $\Theta_\mathrm{j}$ may be correlated with the asymmetry of the particle and thus may yield additional hidden information on sub-diffraction-sized particles.

\section{Conclusion}
The shown straightforward method of spectro-polarimetry with a polarizer and an analyzer in conjunction with the \ac{ADM} is demonstrated to reliably extract the orientations of dipoles excited within asymmetric nanostructures. It is shown that the number of dipole moments is not limited to only transversal and longitudinal \ac{LSP} modes but can be advanced to include dipolar contributions of higher-order modes as well. Although the extraction of \textsmaller{3D} orientations of dipole moments is not performed in this work, it is worth mentioning that the ADM is capable of such analyses. Adjusting the direction of the observer, e.g. by tilting the objective, could resolve the out-of-plane orientation of dipole moments sufficiently for the \ac{ADM} to extract both tilt and azimuthal information. Furthermore, first indications are shown that the method contributes towards spectroscopically determining the azimuthal range of the geometric asymmetry. This is presented for plasmonic nanoparticles by implementing different analyzer angle sets $m$, enabling the creation of large data populations, which are useful for two key points. First, the standard deviation of the delivered dipole orientation populations decreases with rising set number $m$, falling well to the sub-degree level. This does not depend on the number of permutations, but rather on the number of spectra used for the simultaneous fitting procedure, which allows for more reliable ADM fitting. The calculated orientation means converge towards the dipole orientations obtained by fitting all available polarimetric spectra per particle. Control angles taken from \ac{BEM} simulations and \ac{SEM}-extracted highest symmetry axes largely verify the extracted angles.

This makes the presented method suitable for experiments in which the dipole moment orientations of nanostructures need to be extracted without using \ac{SEM} observations, which require vacuum conditions and could permanently alter the sample. The second benefit of a large data population is that the extensive datasets allow the application of \ac{ML} tools to examine the importance values of analyzer angles within various spectral combinations. Observations of the angular spread across all spectra and their permutations could be linked to surface and edge anisotropies due to the degradation of prediction accuracy in the model for variations in individual analyzer angles. The all-spectroscopic polarization method for extracting dipole orientations, combined with the robustness and scalability of the \ac{HGBR} models, strong predictive performance, and insightful analysis, proves valuable for complex or even chiral systems in determining their respective dipole orientations and asymmetry distribution. Fields that involve orientation-specific spectroscopy and are not conducive to \ac{SEM} observations, such as chiroptical and fluorescence spectroscopy, may benefit from this method.

\section{Materials and Methods}
\subsection{Analytical dipole model}\label{subsec:ADM}
The expression for the electric far-field radiation of a dipole is given by Equation~\ref{eq:dip_rad}, which states that the radiation at the point of the observer $\mathbf{r}$ is given by the dipole moment $\mathbf{p}$ as defined in standard textbooks, such as \cite[p.~411]{jackson1999}. The resulting radiation travels in the direction of the observer $\hat{\mathbf{n}}=\mathbf{r}/r$. The electric dipole moment is henceforth referred to as the dipole moment. The general Coulomb force constant, as well as the wavenumber $k$, and distance to the observer $r$, are later included in the general factor $A$. Only the angular frequency $\omega$, time $t$, and phase $\varphi$ are discussed for now. A dipole moment here is defined as $\mathbf{p}_\mathrm{k}=d_\mathrm{k}\parens{\omega} \cdot \mathbf{e}_\mathrm{k}$ with the frequency-dependent magnitude $d_\mathrm{k}\parens{\omega}$ and the orientation vector $\mathbf{e}_\mathrm{k}$.
\begin{equation}\label{eq:dip_rad}
	\mathbf{E}_\mathrm{rad} \approx A \frac{\mathrm{e}^{\parens{\mathrm{i}kr - \mathrm{i}\omega t + \mathrm{i}\varphi}} }{r} \hat{\mathbf{n}} \times \parens{ \hat{\mathbf{n}} \times \mathbf{p}} = A \frac{\mathrm{e}^{\parens{\mathrm{i}kr - \mathrm{i}\omega t + \mathrm{i}\varphi}} }{r} \parens{\hat{\mathbf{n}}\parens{\hat{\mathbf{n}}\cdot\mathbf{p}} - \mathbf{p} }
\end{equation}
Each dipole is being excited by the incident plane wave with the polarization $\mathbf{E}_0$, which scales the far-field radiation of the $k$-th dipole, $\mathbf{E}_\mathrm{k}=(\mathbf{E}_0\cdot\mathbf{e}_\mathrm{k} ) \mathbf{E}_\mathrm{rad}$. For multiple dipoles, the following superposition field is observed:
\begin{equation}\label{key}
	\mathbf{E} \approx \sum_{\mathrm{k}=1}^{\mathrm{N}} \mathbf{E}_\mathrm{k} = A\parens{r}\mathrm{e}^{-\mathrm{i}\omega t} \sum_{\mathrm{k}=1}^{\mathrm{N}}d_\mathrm{k}\parens{\omega}\mathrm{e}^{\mathrm{i}\varphi_\mathrm{k}}\parens{\mathbf{E}_0\cdot\mathbf{e}_\mathrm{k}}\parens{\hat{\mathbf{n}}\parens{\hat{\mathbf{n}}\cdot\mathbf{e}_\mathrm{k}} - \mathbf{e}_\mathrm{k} }
\end{equation}
where the dipole moment $\mathbf{p}$ is substituted by its magnitude $d_\mathrm{k}\parens{\omega}$ and orientation $\mathbf{e}_\mathrm{k}$. The factor $A\parens{r}$ now also includes the radial components $\mathrm{e}^{\mathrm{i}kr}/r$. This orientation-specific electric far field passes through the analyzing filter, which is defined by its transmission axis or parallel axis vector $\mathbf{e}_\mathrm{p}$ given by the orientation angle $\Theta$. The vector $\mathbf{e}_\mathrm{p}$ lies in the plane of the analyzing filter, with the surface normal parallel to the direction of the observer $\hat{\mathbf{n}}$. This scales the electric field component by $\mathbf{E}\cdot\mathbf{e}_\mathrm{p}$. Thus, we define the polarization-dependent electric far field as shown in Equation~\ref{eq:almost_there}, where we can substitute $R_\mathrm{k}  = \mathbf{E}_0\cdot\mathbf{e}_\mathrm{k}$ and $S_\mathrm{k}= (\hat{\mathbf{n}}\cdot \mathbf{e}_\mathrm{p} )(\hat{\mathbf{n}}\cdot\mathbf{e}_\mathrm{k} )-\mathbf{e}_\mathrm{k}\cdot\mathbf{e}_\mathrm{p}$. The factor $R_\mathrm{k}$ expresses the induced scaled dipole moment, and $S_\mathrm{k}$ states the relationship between the dipole $\mathbf{e}_\mathrm{k}$ and the analyzing polarization filter $\mathbf{e}_\mathrm{p}$ regarding the direction of the observer $\hat{\mathbf{n}}$. Looking more closely at $S_\mathrm{k}$ reveals that for complete orthogonality between the incident direction $\mathbf{k}_\mathrm{exc}$ and the analyzer orientation, the factor reduces to $S_\mathrm{k}= -\mathbf{e}_\mathrm{k}\cdot\mathbf{e}_\mathrm{p}$, thus just stating the scaling of the dipole moment.
\begin{equation}\label{eq:almost_there}
	\begin{split}
		E_\mathrm{p} &= A\parens{r}\mathrm{e}^{-\mathrm{i}\omega t} \sum_{\mathrm{k}=1}^{\mathrm{N}}d_\mathrm{k}\parens{\omega}\mathrm{e}^{\mathrm{i}\varphi_\mathrm{k}}\parens{\mathbf{E}_0\cdot\mathbf{e}_\mathrm{k}}\parens{\parens{\hat{\mathbf{n}}\cdot\mathbf{e_\mathrm{p}}}\parens{\hat{\mathbf{n}}\cdot\mathbf{e}_\mathrm{k}} - \mathbf{e}_\mathrm{k}\cdot\mathbf{e}_\mathrm{p}}  \\
		&= A\parens{r}\mathrm{e}^{-\mathrm{i}\omega t} \sum_{\mathrm{k}=1}^{\mathrm{N}}d_\mathrm{k}\parens{\omega}\mathrm{e}^{\mathrm{i}\varphi_\mathrm{k}}R_\mathrm{k}S_\mathrm{k}
\end{split}
\end{equation}
The total intensity from a collection of $N$ independent dipoles can then be evaluated using Equation~\ref{eq:intensity}, which expresses the frequency-dependent intensity as a function of the dipole magnitudes, their corresponding phase responses and orientations, the illumination orientation and polarization, and the orientation of the analyzing filter.
\begin{equation}\label{eq:intensity}
\begin{split}
	I_\mathrm{p}\parens{\omega} &\approx \Big| E_\mathrm{p} \Big|^2 = E_p \cdot E_p^{*} =\\
	&= \Big| A\parens{r} \Big|^2 \brackets{ \sum_{\mathrm{k}=1}^{\mathrm{N}} d_\mathrm{k}\parens{\omega}R_\mathrm{k}^2S_\mathrm{k}^2 + 2 \sum_{\mathrm{k}=1}^{\mathrm{N}} \sum_{\mathrm{l}>\mathrm{k}}^{\mathrm{N}} d_\mathrm{k}\parens{\omega}d_\mathrm{l}\parens{\omega} R_\mathrm{k}R_\mathrm{l}S_\mathrm{k}S_\mathrm{l}\cdot\cos{\parens{\varphi_\mathrm{k}-\varphi_\mathrm{l}}} } 
\end{split}
\end{equation}

\begin{figure*}[t!]
	\centering
	\includegraphics[width=\textwidth]{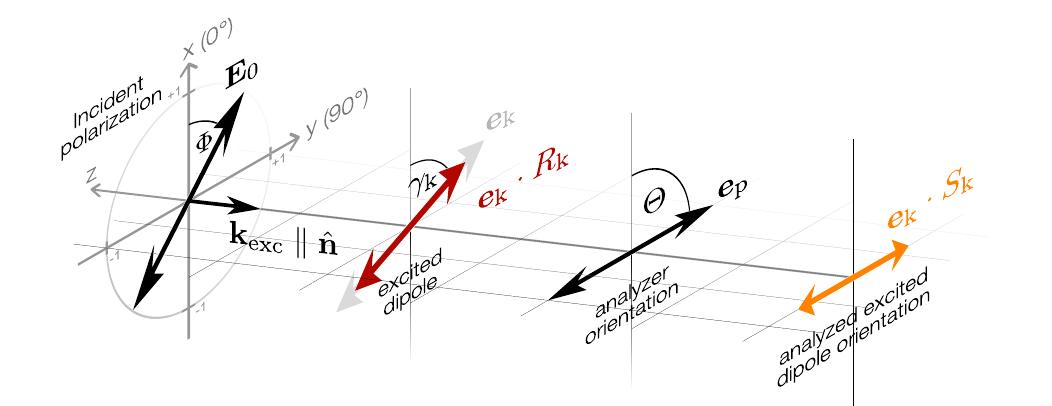}
	\caption{Illustration of the \ac{ADM} considering both incident analysis polarizations. The incident plane wave with the polarization of $\mathbf{E}_0$ with angle $\Phi$ with respect to the \textit{x}-axis travels in the negative \textit{z}-direction. The wave vector $\mathbf{k}_\mathrm{exc}$ is therefore parallel to the direction of the observer $\hat{\textbf{n}}$. The dipole with orientation $\mathbf{e}_\mathrm{k}$ is excited through the incident polarization and thus scaled with $R_\mathrm{k}$. As the signal passes through the analyzer, its orientation $\mathbf{e}_\mathrm{p}$ will reveal the component of the excited dipole that is parallel to $\mathbf{e}_\mathrm{p}$, resulting in scaling with the dot product $S_\mathrm{k}$ (see Equation~\ref{eq:almost_there}).}\label{fig:ADM}
\end{figure*}

By looking at the normal incidence case, the excitation and propagation are reduced to the negative \textit{z}-direction (see Figure~\ref{fig:ADM}). Consequently, the investigation is restricted to \textit{x-y} projections of the dipole orientations, which allows us to evaluate the in-plane dipoles of planar particles. Their respective orientation vector can be modeled with simple trigonometry, e.g. $\mathbf{e}_\mathrm{k} = \cos{\parens{\gamma_\mathrm{k}}}\cdot\hat{\mathbf{e}}_\mathrm{x} + \sin{\parens{\gamma_\mathrm{k}}}\cdot\hat{\mathbf{e}}_\mathrm{y}$, with $\gamma_\mathrm{k}$ representing the azimuthal angle of the $k$-th dipole. Similarly, the excitation polarization and far-field analyzation azimuthal angles $\Phi$ and $\Theta$ assign the \textit{x-y} angle of the optical elements. Abiding by the standard mathematical convention, the azimuthal angle starts at 0° from the positive \textit{x}-axis direction and turns counterclockwise around the \textit{z}-axis with increasing angular value. Lastly, by specifying the number of observed dipoles and their spectral magnitudes and phase responses, Equation~\ref{eq:intensity} can be used to express the spectral features as a function of the analyzer orientation angle, thereby resolving the recorded intensity of the dipole interplay \cite{mildner2023, simo2025}. The spectral magnitudes $d_\mathrm{k}\parens{\omega}$ are calculated with a skewed Lorentzian, from which the \ac{FWHM}, resonance $\lambda_\mathrm{k}$ and skewness $g$ and the phase responses $\varphi_\mathrm{k}\parens{\omega}$ can be calculated. 

Experimentally, the magnitudes can be extracted from an unpolarized spectrum of the system fitted with a set of skewed Lorentzians, from which the individual phase responses can then be calculated. For the polarimetric measurements, the experiment’s fixed excitation polarization angle $\Phi$ and far-field analyzation angles $\Theta$ are known. Since the planar nanostructures explored here will mainly express two dipoles, Equation~\ref{eq:intensity} can be reduced to exemplary two dipoles with in-plane angles $\gamma_1$ and $\gamma_2$, with the respective magnitudes $d_1\parens{\omega}$ and $d_2\parens{\omega}$ and phase responses $\varphi_1\parens{\omega}$ and $\varphi_2\parens{\omega}$. Equation~\ref{eq:ADM} is referred to as the \acf{ADM} throughout this paper \cite{simo2025}.

\begin{equation}\label{eq:ADM}
\begin{split}
	I_\mathrm{p}\parens{\omega,\Theta} = A^2\big[ &d_1\parens{\omega}\cos^2\parens{\Phi-\gamma_1}\cos^2\parens{\Theta-\gamma_1} \\ +&d_2\parens{\omega}\cos^2\parens{\Phi-\gamma_2}\cos^2\parens{\Theta-\gamma_2} \\
	+&2d_1\parens{\omega}d_2\parens{\omega}\cos\parens{\Phi-\gamma_1}\cos\parens{\Phi-\gamma_2}\\
	\cdot & \cos\parens{\Theta-\gamma_1} \cos\parens{\Theta-\gamma_2} \cos\parens{\varphi_1\parens{\omega} - \varphi_2\parens{\omega}}\big]
\end{split}
\end{equation}

The first two terms express the modulation of the respective dipole intensity, originating from the dot products of $R$ and $S$ as stated in Equation~\ref{eq:almost_there}. The third term, however, includes the influence of each magnitude with respect to the relative phase response between the two dipoles. Thus, we can see how the dipole intensities influence one another, depending on their properties, as revealed by the analysis of the far field (see Figure~\ref{fig:ADM}). Naturally, the complexity of the \ac{ADM} increases with the number of observable dipoles. The expression for the \ac{ADM} for three dipoles is available in the Supporting Information (see Equation S1). These models are used for the dipole moment orientation analysis, in accordance with the number of modes expressed in the presented asymmetric nanostructures.

\subsection{Fabrication}\label{subsec:fabrication}
Asymmetric plasmonic gold nanorods are synthesized in-house through a seed-mediated method, following the recipe provided by Scarabelli \textit{et al.} to nominally produce gold nanorods \cite{scarabelliTipsTricksPractical2015}. The synthesis consisted of first producing a single-crystal gold seed suspension by fast injection of sodium borohydride (\ch{NaBH4}) in an aqueous solution of hydrogen tetrachloroaurate (\ch{HAuCl4}) and hexadecyltrimethylammonium bromide (\ch{CTAB}). A small volume of seed suspension is added to a growth solution, which also consists of \ch{HAuCl4} and \ch{CTAB}. The pH of the solution is adjusted to 1.5 with hydrochloric acid (\ch{HCl}). Ascorbic acid as a reduction agent and silver nitrate (\ch{AgNO3}) as the anisotropic growth mediator are added to the growth solution before a small volume of the seed suspension is introduced. To increase the nanoparticle asymmetry, the synthesis is altered by extending growth times beyond the recommended time of 2\,h and by varying the volume of \ch{AgNO3} in the growth solution. The particles in Figure~\ref{fig:specimens}(a,b) appear to each consist of several smaller nanorods that are grown together. The size distribution in the resulting suspension lies at approximately 100\,nm $\pm$ 25\,nm in the longitudinal axis. The asymmetric nanorods are spin-coated onto a glass substrate coated with 50\,nm of conductive \acf{ITO}. The glass/\ac{ITO} substrate is also pre-structured with a marker grid, to make correlative \ac{SEM} observations possible after the optical characterization. A third nanostructure is fabricated through \ac{EBL} and a lift-off process \cite{chenNanofabricationElectronBeam2015}. It is nominally designed as a 20\,nm thick gold rectangle with dimensions of 100\,nm $\times$ 50\,nm in \textit{x}- and \textit{y}-direction, respectively. The structure’s asymmetry is realized by overdevelopment of the resist layer.

\subsection{Polarimetric measurements}\label{subsec:measurements}
All plasmonic systems are first analyzed through dark-field transmission microscopy under unpolarized excitation with a broadband white-light halogen lamp, followed by polarimetric measurements. An inverted microscope (Eclipse Ti-S, \textit{Nikon}) is equipped with a dry \ac{DFC} (NA\,=\,0.8 - 0.95, \textit{Nikon}). An ultra-broadband linear polarizing filter (WP25M-UB, \textit{Thorlabs}) is inserted between the light source and the \ac{DFC}. Additionally, the depolarizing components of the \ac{DFC} are filtered out with a cross aperture at the \ac{DFC} annulus, allowing only the s- and p- components of the incident light to reach the plasmonic particle \cite{simo2025, chenObservationFanoResonance2011}. The dark-field polarimetric measurements are performed with incident linear polarization fixed at 45° to the \textit{x}-axis of the piezo scanning stage (P-545.xR7, \textit{Physik Instrumente}) (see Supporting Information, Figure S1). The dark-field scattering signal is collected in transmission with a 100$\times$ oil objective (CFI Plan Fluor, \textit{Nikon}) adjusted to an NA\,=\,0.75. Directly after the objective, a piezoelectric rotational mount (ELLK14, \textit{Thorlabs}) equipped with another ultra-broadband polarizing filter (WP25M-UB, \textit{Thorlabs}) analyzes the collected scattered light in 15° increments from 0° to 165°, which is then recorded by a spectrograph (Shamrock SR-303i, \textit{Andor}) equipped with a \ac{CCD} detector (iDus 416 A-LDC-DD, \textit{Andor}). The \ac{CCD} detector is cooled down to --\,60\,°C to reduce dark-current noise. The measurements are background-corrected for each analyzer position. The complete set of measurements is subsequently normalized to the highest intensity recorded within the set for further analysis. Afterward, the plasmonic particles are observed in the \ac{SEM} at normal incidence to extract their asymmetric contours and the angles of the longest and shortest geometric symmetry axes, $\gamma_\mathrm{long}$ and $\gamma_\mathrm{short}$, respectively.

\subsection{Contour extraction and boundary element method (BEM) simulations}\label{subsec:BEM}

The SEM micrograph of a particle provides a \textsmaller{2D} projection of a plasmonic particle. Its contour is extracted with the OpenCV package in Python, where the \ac{SEM} micrograph is used to gain equidistant pixel positions from the grayscale image \cite{OpenCV_itseez2015}. The asymmetric contour is then used to extrude and mesh the \textsmaller{3D} model for the \ac{BEM} simulation, performed with the Matlab package MNPBEM \cite{MNPBEM_waxenegger2015, MATLAB_PDE2024}. The particle model is set on a layered structure, identical to the experiment’s substrate, featuring 50\,nm \ac{ITO} on glass. The refractive index data for \ac{ITO} has been evaluated in-house through ellipsometry \cite{sommer2024}. The extrusion was locally weighted on the \textsmaller{3D} model's surface, such that it imitates the grayscale intensity as a height profile. For simplicity, the conversion of the grayscale values to particle heights is assumed as a linear function. The \textsmaller{3D} extrusion slope for each system is adjusted to best fit the resonance wavelengths of the experimental unpolarized dark-field scattering spectrum. It is evident that the extrusion of the plasmonic particle contour only approximates the actual morphology of the particle. Still, the extrusion height helps to mimic the experimental scattering data, while the \textit{x-y} projection is of higher relevance for extracting the dipole moment orientation in the same plane. The simulations calculate the surface charge densities $\rho\parens{\textbf{r}',t}$, which deliver the time-harmonic dipole moment $\mathbf{p}_\mathrm{k,eig}\parens{t}\in C^{3\times1}$ of the corresponding \ac{LSP} mode $k$ at the resonance frequency by calculating $\mathbf{p}_\mathrm{k,eig}\parens{t} = \sum_{i} r'_i\cdot \rho_i \parens{r',t} \cdot\Delta S'_i$, where $\Delta S'$ are the discretized surface areas at the positions $\mathbf{r}'$ of the nanostructure model \cite[p.~410]{jackson1999}\cite{razimanDoesRealPart2016}. By following the instructions of Dennis \textit{et al.} \cite{dennisPolarizationSingularitiesParaxial2002} we can extract the simulated dipole moments at the correct phase. The cited procedure originally calculates the phase that corresponds to the time-averaged electric field. This is named the rectifying phase; the real and imaginary parts of the electric field are orthogonal and aligned to the principal axes of the polarization ellipse. The present work applies this approach to the dipole moments, as they are directly related to the electric field. For more details refer to reference \cite{dennisPolarizationSingularitiesParaxial2002}, as the following explanation uses the same notation. The time-harmonic dipole moment traces an ellipse with major and minor axes $\mathbf{a}$ and $\mathbf{b}$. The traced ellipse can be expressed as the equation $\mathbf{a}+\mathrm{i}\mathbf{b} = \mathbf{p}_\mathrm{k,eig}\parens{t}\cdot\mathrm{e}^{-\mathrm{i}\chi}$, with $\chi=\omega t$. At the rectifying phase $\chi_0$, the real and imaginary parts of $\mathbf{p}_\mathrm{k,eig}\parens{t}$, denoted with $\mathbf{p}_\mathrm{Re}$ and $\mathbf{p}_\mathrm{Im}$, are orthogonal and coinciding with the major and minor axis of the traced ellipse. Ultimately, we can express the time-averaged dipole moment, and consequently $\chi_0$, as follows: 

\begin{align}
	\mathbf{p}_\mathrm{k,eig}\cdot\mathrm{e}^{-\mathrm{i}\chi_0} &= \mathbf{p}_\mathrm{Re}\cos\parens{\chi_0} + \mathbf{p}_\mathrm{Im}\sin\parens{\chi_0} + \mathrm{i}\brackets{\mathbf{p}_\mathrm{Im}\cos\parens{\chi_0} -\mathbf{p}_\mathrm{Re}\sin\parens{\chi_0}}=\mathbf{a}+\mathrm{i}\mathbf{b} \label{eq:polarization_ellipse}\\
	\chi_0 &= \frac{1}{2}\arctan\parens{\frac{2\mathbf{p}_\mathrm{Re}\cdot\mathbf{p}_\mathrm{Im}}{p^2_\mathrm{Re}-p^2_\mathrm{Im}}} \label{eq:rectifying_phase}
\end{align}

The ellipse's azimuthal angle $\gamma_\mathrm{k,eig}$ between the \textit{x}-axis and its major axis a can be generally calculated by $\gamma_\mathrm{k,eig} = \arctan\parens{a_\mathrm{y}/a_\mathrm{x}}$. Using Equation~\ref{eq:polarization_ellipse} and Equation~\ref{eq:rectifying_phase}, one can write $\gamma_\mathrm{k,eig}$ in terms of the real and imaginary components of the simulated dipole moment (see Equation~\ref{eq:dipole_orientation}).

\begin{equation}\label{eq:dipole_orientation}
	\gamma_\mathrm{k,eig} = \frac{1}{2}\arctan\parens{\frac{2p_\mathrm{Re,x} p_\mathrm{Im,x} + 2 p_\mathrm{Re,y}p_\mathrm{Im,y}}{p^2_\mathrm{Re,x} + p^2_\mathrm{Re,y} - p^2_\mathrm{Im,x} - p^2_\mathrm{Im,y}}}
\end{equation}

\subsection{Machine learning model}
As the \ac{ML} algorithm, an \ac{HGBR} from the Scikit-learn Python library version 1.2.0 is chosen, which is an ensemble \ac{ML} algorithm based on gradient boosting methods with histogram-based binning. These properties make this model a prime candidate for large datasets, since it reduces memory consumption with its binning approach and minimizes error through iterative addition of decision trees for corrections of residual error \cite{pedregosaScikitlearnMachineLearning, mayerMachineLearningApplications2022, mansooriMachineLearningBasedPredictive2025, soltanimoghadamEarthquakeLocationMagnitude2024}. It has proven to be a superior model for many research fields and applications handling tabular feature data linked to target data, similar to the filtered permutations of analyzer angles and their respective dipole orientation, as opposed to temporal or spatial data, such as audio and image data \cite{davisComparisonGaitSpeed2021, lundbergLocalExplanationsGlobal2020, chenXGBoostScalableTree2016}. 
The model is trained with the filtered permutations of analyzer angles as its features and the resulting orientations per dipole as targets. The features are given as $\Theta_\mathrm{j}\in\brackets{\Theta_1,\Theta_2,\Theta_3,\Theta_4,\Theta_5,\Theta_6}$, where one such row displays one permutation of analyzer angles that delivers one target value, an orientation angle. A fresh \ac{HGBR} model is created and optimized for hyperparameters such as regularization, learning rate, maximum depth, and maximum iterations with the grid search \ac{CV} method included in the Scikit-learn package to ensure high accuracy. The dataset is split into training and testing subsets with an 80:20 ratio, respectively. Hyperparameter investigation and model training use the training data subset, after which the model’s orientation predictions are compared to the testing data. A sub-degree discrepancy in \acf{RMSE} and a low percentage discrepancy in R\textsuperscript{2} score are evaluated, which shows that no overfitting is present in the model. Overfitting of the model to the training data subset is a common pitfall in \ac{ML}, which is why this is thoroughly investigated. The model’s capabilities are ultimately validated by using the whole dataset for testing, where a 5-fold \ac{CV} reaches metrics such as R²\,$>$\,0.95 and \ac{RMSE}\,$<$\,0.4°, ensuring stable prediction performance (see Table S1 in Supporting Information).

\section*{Acknowledgments}
The authors gratefully acknowledge contributions to the spectroscopy software and model of local polarization after the dark-field condenser by S. Dickreuter. Special thanks is given to Jorge Olimos-Trigo for insightful discussions on polarimetry involving asymmetric particles. Support from the Open Access Publication Fund of the University of Tübingen is gratefully acknowledged. 

\section*{Author contributions}
\textbf{P. Christian Simo}: conceptualization: (supporting); data curation: (lead); formal analysis: (lead); investigation: (lead); methodology: (equal); software: (lead); validation: (lead); writing - original draft: (lead). \textbf{Michaela Zbytovska}: data curation: (supporting); formal analysis: (supporting); investigation: (supporting); writing - review \& editing: (supporting). \textbf{Annika Mildner}: conceptualization: (lead); writing - review \& editing: (supporting). \textbf{Lukas Lang}: software: (supporting); validation: (supporting), writing - review \& editing: (supporting). \textbf{Melanie Sommer}: software: (supporting); validation: (supporting), writing - review \& editing: (supporting). \textbf{Dieter P. Kern}: conceptualization: (lead); methodology: (supporting); validation: (equal); writing - review \& editing: (supporting). \textbf{Monika Fleischer}: project administration: (lead); resources: (supporting); supervision: (lead); validation: (equal); writing - review \& editing: (equal).

\section*{Code and data availability}
The data that support the findings of this study are available from the corresponding author upon reasonable request.

%%%%% References %%%%%
\printbibliography

\end{document}

\maketitle
	
\section{Analytical dipole model for three dipoles}

\begin{equation}
	\begin{split}\label{eq:appendix_ADM_3dip}
		I_{\mathrm{p}}\parens{\omega,\Theta} =\\ 
		A^2 \cdot \big[ & d_1^2\parens{\omega} \cos^{2}{\parens{\Phi - \gamma_1}} \cos^{2}{\parens{\Theta - \gamma_1}}\\
		+ & d_2^2\parens{\omega} \cos^{2}{\parens{\Phi - \gamma_2}} \cos^{2}{\parens{\Theta - \gamma_2}}\\
		+ & d_3^2\parens{\omega} \cos^{2}{\parens{\Phi - \gamma_3}} \cos^{2}{\parens{\Theta - \gamma_3}}\\
		+ & 2 d_1\parens{\omega} d_2\parens{\omega} \cos{ \parens{\Phi - \gamma_1} } \cos{ \parens{\Phi - \gamma_2}} \\
		\cdot & \cos{ \parens{\Theta - \gamma_1} } \cos{ \parens{\Theta - \gamma_2} } \cos{ \parens{\varphi_1\parens{\omega} - \varphi_2\parens{\omega}} } \\
		+ & 2 d_1\parens{\omega} d_3\parens{\omega} \cos{\parens{\Phi - \gamma_1}} \cos{\parens{\Phi - \gamma_3}} \\ 
		\cdot & \cos{\parens{\Theta - \gamma_1}} \cos{\parens{\Theta - \gamma_3}} \cos{\parens{\varphi_1\parens{\omega} - \varphi_3\parens{\omega}}} \\
		+ & 2 d_2\parens{\omega} d_3\parens{\omega} \cos{\parens{\Phi - \gamma_2}} \cos{\parens{\Phi - \gamma_3}} \\ \cdot & \cos{\parens{\Theta - \gamma_2}} \cos{\parens{\Theta - \gamma_3}} \cos{\parens{\varphi_2\parens{\omega} - \varphi_3\parens{\omega}}} \big]
	\end{split}
\end{equation}
	
\section{Optical setup and alignment}

\begin{figure}[t!]
	\centering
	\includegraphics{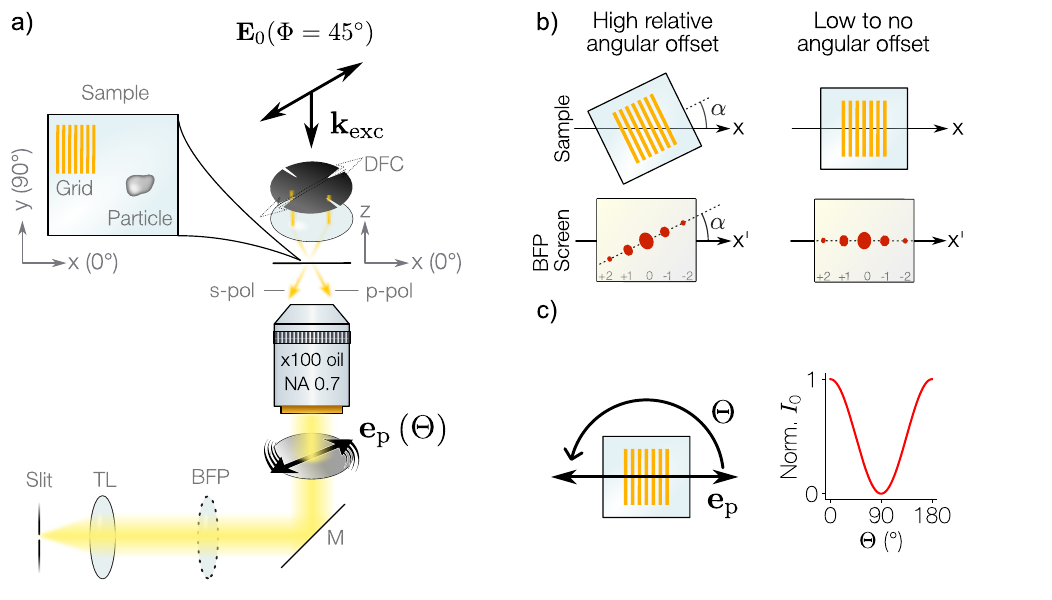}
	\caption{\textbf{(a)} General setup for dark-field polarimetric spectroscopy. The incident linearly polarized light locked at 45° is passed through a cross-aperture in the \ac{DFC}, which allows only s- or p-polarized illumination in the sample plane, removing any depolarizing components. The light is focused on the surface of the substrate, which harbors the plasmonic system, including a \textsmaller{1D} gold grid for orientation alignment purposes (see inset). The collected scattering of the induced dipole moments $\mathbf{p}_\mathrm{k}$ is analyzed by a polarizing filter, which is controlled by a piezo-rotational mount. The orientation of the analyzer is denoted by $\mathbf{e}_\mathrm{p}\parens{\Theta}$. The scattered light is then guided to the slit of the spectrometer with a mirror (M) and a tube lens (TL). A \ac{BFP} lens can be inserted in the pathway for (b) sample and (c) polarization alignment purposes. \textbf{(b)} The sample is aligned with the aid of the on-substrate gold grid. With brightfield illumination, a 10\,nm bandpass filter at 700\,nm, and the \ac{BFP} lens, the grating modes of the grid can be displayed on the CCD camera. The skew of the sample can be adjusted such that the relative offset of the sample is minimized. \textbf{(c)} The orientation of the analyzing filter is varied from 0° to 180° and the respective intensity is recorded. The minimum of the normalized intensity indicates the angle when $\mathbf{e}_\mathrm{p}$ is parallel to the grid.}\label{fig:optical_setup}
\end{figure}

Figure~\ref{fig:optical_setup} displays the inverse dark-field setup equipped with two linear polarizing filters. These define the linearly polarized $\mathbf{E}_0$ incident light and the axis of transmission $\mathbf{e}_\mathrm{p}$ of the analyzing filter. The alignment relative to the sample is achieved by considering a \textsmaller{1D} grating additionally fabricated on the substrates beforehand (see Figure~\ref{fig:optical_setup}). Through Fourier optics, the rotation of the Fraunhofer diffraction pattern of the grating and thus the rotation of the sample can be closely monitored \cite{bonodDiffractionGratingsPrinciples2016}. Polarization filters and \ac{SEM} images are aligned to the gratings, therefore eliminating the relative angle between the observed \ac{SEM} micrograph and the orientation of the samples during spectroscopy.

\section{Collection of dipole orientations from sets of analyzer angles $m$}
\begin{figure}[h]
	\centering
	\includegraphics{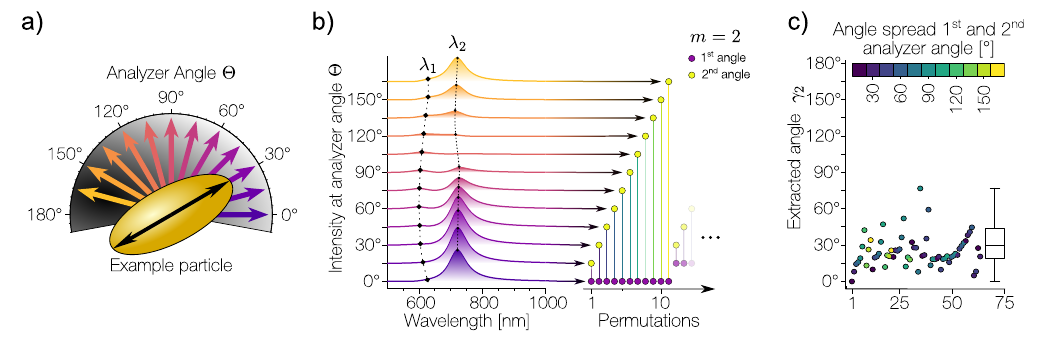}
	\caption{Exemplary iteration process for $m$\,=\,2, meaning a set of two analyzer angles. \textbf{(a)} The particle is polarimetrically measured with the set analyzer angles. Its longitudinal dipole moment is denoted with the black arrow. \textbf{(b)} Each analyzer angle delivers a spectrum with two resonance wavelengths $\lambda_1$ and $\lambda_2$, marking the transversal and longitudinal resonance modes in this case. The polarimetric spectra are paired up in sets of two. The first combination is composed of the spectrum taken with the analyzer angle at 0° paired with the spectrum taken at 15° analyzer orientation. These pairs are fitted simultaneously with the \ac{ADM}, which delivers angles for the two desired dipole orientations. This is done iteratively for all possible combinations. \textbf{(c)} All combinations deliver extracted angles $\gamma_2$ each, which are collected and evaluated statistically. Extracted angle values outside the \ac{IQR} and the respective angle combinations they resulted from are filtered and removed. Here, the extracted angles $\gamma_2$ are additionally labeled with colors, which represent the fanning between the two analyzer angles. }\label{fig:combinations}
\end{figure}

The iteration process for two simultaneously fitted analyzer angles ($m$\,=\,2), which have a total of $C_{m=2}^{n=12}$\,=\,66 possible combinations within the given set of analyzer angle positions ([0°, 15°, 30°, 45°, 60°, 75°, 90°, 105°, 135°, 150°, 165°]) (see Figure~\ref{fig:combinations}a), can be pictured as follows: The first analyzer angle is static at 0°, while the second analyzer angle travels from 15° to 165° (see Figure~\ref{fig:combinations}b). Then the first analyzer angle is switched to 15°, while the second analyzer angle travels from 30° to 165°. This process continues until the first analyzer angle reaches 150° and the second analyzer angle is at 165°. Each combination delivers one dipole moment orientation $\gamma_\mathrm{k}$ for each of the $k$ modes.  Each pool of values is firstly evaluated with a box plot, to extract statistical properties such as the first quartile $Q_1$, the second quartile $Q_2$ also known as the median, the third quartile $Q_3$ and the interquartile region $\mathrm{IQR}=Q_3-Q_1$. Upper and lower boundaries were set with $Q_1-1.5\times\mathrm{IQR}$ and $Q_3+1.5\times\mathrm{IQR}$ to remove any extreme outliers, reducing the total combinations $C$ to $\tilde{C}$. If the minimum and/or maximum value of the set lie within these boundaries, then the respective boundary is set to the corresponding value. The simultaneous fitting and establishment of all possible iterations and computations is orchestrated using mainly the SciPy Python library version 1.15.3 and other standard Python packages \cite{virtanenSciPy10Fundamental2020,scipy_vanrossum2020}.
	
\section{Histogram gradient boosting regressor errors per particle}

\begin{table}[h]
	\centering
	\caption{The \ac{RMSE} in degrees and the R\textsuperscript{2} scores of the \ac{HGBR} model trained on the dipole orientations $\gamma_\mathrm{k}$ for the set $m$\,=\,6. Each particle was evaluated separately, which delivered sub-degree residual errors and high R\textsuperscript{2} values.}\label{tab:errors_ml}
	\begin{tabular}{r|c | c | c}
		Particle & Low asymmetry & High asymmetry & EBL\\ [0.5em]
		\hline\hline
		\ac{RMSE} [°] & 0.321 & 0.293 & 0.393 \\ [0.5em]
		R$^2$ & 0.976 & 0.984 & 0.978 \\ [0.5em]
		
	\end{tabular}
\end{table}

\section{Scalar decomposition of surface charge density distribution}
\begin{figure}[h]
	\centering
	\includegraphics[width=\textwidth]{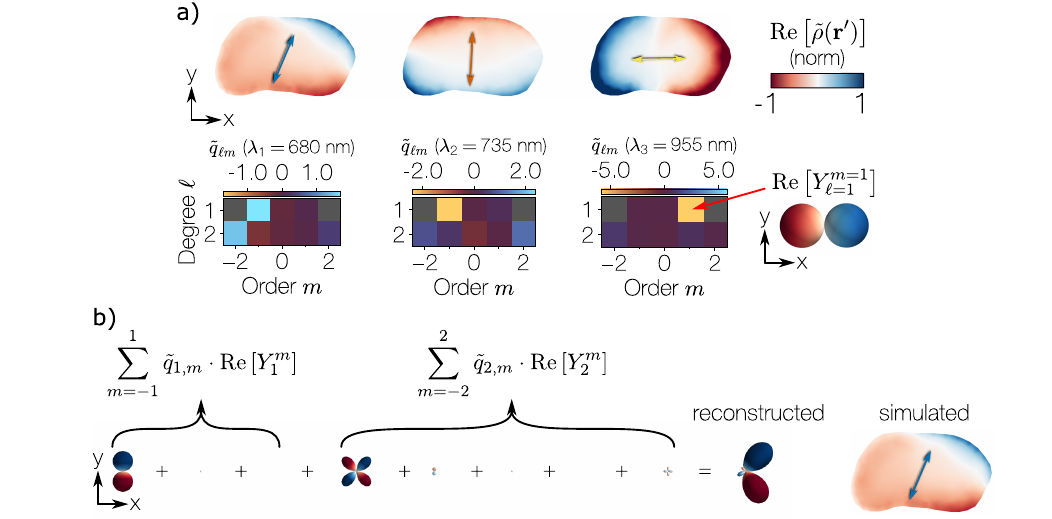}
	\caption{\textbf{(a)} Multipole decomposition of the surface charges $\tilde{\rho}\parens{\mathbf{r}'}$ of the \ac{EBL}-fabricated particle at the \ac{LSP} mode wavelengths $\lambda_\mathrm{k}$, cf. Figure 3c. The magnitudes of the multipole moments $\tilde{q}_{\ell m}$ are given for each degree $\ell$ and order $m$. A higher magnitude signifies a higher weight of the respective spherical harmonic $Y_\ell^m$, and the sign describes whether $Y_\ell^m$ is in-phase (positive) or out-of-phase (negative). An exemplary $Y_\ell^m$ is given for $\ell$\,=\,1 and $m$\,=\,1, which is out-of-phase for the surface charges of the longitudinal mode at $\lambda_3$\,=\,955\,nm. \textbf{(b)} Recomposition of the surface charges at $\lambda_1$\,=\,680\,nm by summing the respective spherical harmonics with their weight coefficients. The dipolar degree $\ell$\,=\,1 adds the highest contributions, followed by the quadrupole. The resulting multipole resembles the charge density distributions, with a clearly dipolar characteristic.}\label{fig:scalar_decomposition}
\end{figure}

The multipole decomposition was performed on the simulated surface charge densities $\tilde{\rho}\parens{\mathbf{r}'}$ at the positions of the surface elements $\mathbf{r}'$ with areas $dS'$. The multipole weights or multipole coefficients $\tilde{q}_{\ell m}$ are calculated with the complex conjugates of the complex spherical harmonics $Y_\ell^{m}\parens{\theta',\phi'}$ at the angles $\theta'$ and $\phi'$ of the surface elements $dS'$ (see Equation~\ref{eq:weights}).
\begin{equation}\label{eq:weights}
	q_{\ell m} = \sum_{i}^{\ell_\mathrm{max}} r_i'^\ell\cdot\tilde{\rho}_i\parens{\mathbf{r}_i'}\cdot Y_\ell^{m}\parens{\theta',\phi'}~dS'
\end{equation}
Iterating the process for the degrees $\ell\in\brackets{1,2}$ with the orders $m\in\brackets{-\ell,\ell}$, encompassing the dipolar and quadrupolar multipoles, extracts their respective weights $q_{\ell m}$. Due to the symmetry condition of the spherical harmonics, which is reliant on the sign of the order $m$, we extract $\tilde{q}_{\ell m}$ as follows
\begin{equation}
	\tilde{q}_{\ell m} = \left\{ 
	\begin{array}{ll}
		(-1)^{m} \sqrt{2} \, \mathrm{Im}\brackets{q_{\ell m}}, & \text{for } m < 0 \\[1ex]
		\hfill \mathrm{Re}\brackets{q_{\ell m}}, & \text{for } m = 0 \\[1ex]
		\hfill \sqrt{2} \, \mathrm{Re}\brackets{q_{\ell m}}, & \text{for } m > 0
	\end{array} 
	\right.
\end{equation}
These coefficients portray the weight and phase of the spherical harmonics through their value and sign, respectively. They allow us to qualitatively describe the multipole characteristics of scalar values, such as the surface charge densities, and therefore describe their distributions (see Figure~\ref{fig:scalar_decomposition}a for the \ac{EBL}-fabricated particle). Looking at the contributions of the dipole ($\ell$\,=1) shows the \textit{x}- and \textit{y}-type dipole ($m$\,=\,$\pm$1) to be the leading actors in the surface charge distribution, supporting the fact that the orientation of the dipole is mainly in the plane of the \ac{EBL} structure at the \ac{LSP} mode resonance wavelengths $\lambda_\mathrm{k}$. The reconstruction of the surface charge distributions for $\lambda_1$ reveals that the mode contains \textit{y}-dipole contributions and an in-plane quadrupolar contribution of lower weight, which results in an asymmetric but largely dipolar distribution (see Figure~\ref{fig:scalar_decomposition}b). This closely resembles the actual simulated surface charge distribution.

%%%%% References %%%%%
\printbibliography